\documentclass[a4paper,11pt]{article}
\pdfoutput=1

\usepackage{amsmath}
\usepackage{graphicx}
\usepackage{jheppub}
\usepackage{xspace}
\usepackage{booktabs}

\newcommand{\MS}{\ensuremath{\overline{\text{MS}}}\xspace}
\newcommand{\centeredgraphics}[2][]{\vcenter{\hbox{\includegraphics[#1]{#2}}}}
\DeclareMathOperator{\tr}{tr}
\DeclareMathOperator{\str}{str}
\DeclareMathOperator{\Li}{Li}

\hypersetup{
  pdftitle={Quark Condensate at Five Loops},
}

\graphicspath{{figures/}{build/figures/}}

\title{%
  Heavy-Quark Condensate and Vacuum Energy Anomalous Dimension at Five Loops
}

\author[a,b]{Andreas Maier}
\emailAdd{andreas.maier@ifj.edu.pl}
\author[c]{Peter Marquard}
\emailAdd{peter.marquard@desy.de}
\author[d]{York Schröder}
\emailAdd{yschroder@ubiobio.cl}
\affiliation[a]{%
The Henryk Niewodniczański Institute of Nuclear Physics,
ul. Radzikowskiego 152, 31-342 Krakow, Poland
}
\affiliation[b]{%
  Institut de F{\'i}sica d'Altes Energies (IFAE),
  The Barcelona Institute of Science and Technology,
  Campus UAB, 08193 Bellaterra (Barcelona), Spain
}
\affiliation[c]{%
  Deutsches Elektronen-Synchrotron DESY, Platanenallee 6, 15738 Zeuthen, Germany
}
\affiliation[d]{%
  Centro de Ciencias Exactas, Departamento de Ciencias Básicas,
  Universidad del Bío-Bío, Avenida Andrés Bello 720, Chillán, Chile
}

\abstract{%
  We present the perturbative heavy-quark condensate at five-loop
  order in QCD. This constitutes the first calculation at this order
  accounting for non-vanishing quark masses. Our result confirms the
  computation of the five-loop vacuum anomalous dimension by Baikov and
  Chetyrkin.
}

\preprint{DESY-26-047}

\begin{document}
\maketitle

\section{Introduction}
\label{sec:intro}

The quark condensate $\langle \bar{q} q\rangle$ plays a central role
in the dynamics of the strong interactions. Its nonvanishing value in
the limit of massless quarks, $m_q\to 0$, signals the spontaneous
breaking of chiral symmetry. For physical up- and down-quark masses
$m_u$ and $m_d$, the pion can be identified as the associated pseudo
Goldstone boson. As shown by Gell-Mann, Oakes, and
Renner~\cite{Gell-Mann:1968hlm}, the mass $M_\pi$ and decay constant
$F_\pi$ of the pion are closely related to the up-quark condensate
$\langle \bar{u} u \rangle$ via
\begin{equation}
  \label{eq:GMOR}
  M_\pi^2 F_\pi^2 = (m_u + m_d) \langle \bar{u} u \rangle + \mathcal{O}(m_q^2).
\end{equation}
In practice, the value of the quark condensate is often required in
applications of the operator product expansion (OPE)
\cite{Wilson:1969zs}. Indeed, it is the leading non-trivial local
operator contributing to the short-distance expansion of a
gauge-invariant operator product $\mathcal{O}(x) \mathcal{O}(0)$ with
odd mass dimension,\footnote{For even-dimensional products the gluon
condensate $\langle G^2 \rangle$ contributes at the same expansion
order as the quark condensate.}
\begin{equation}
  \label{eq:OPE}
  \langle \mathcal{O}(x) \mathcal{O}(0) \rangle = C_1(x) + C_{\bar{q} q}(x) \langle \bar{q} q \rangle + \dots,
\end{equation}
where the ellipsis indicates higher-order terms in the expansion
around $x=0$. Applied to the quark self-energy, this feature of the
OPE suggests an interpretation of the quark condensate as an effective
long-distance quark mass~\cite{Politzer:1976tv}.

In many cases, the OPE is used at low energies, where only the
non-perturbative light-quark condensates are relevant. Conversely, at
sufficiently high energies non-perturbative effects become negligible
and only perturbative heavy-quark condensates need to be considered.
In this scenario, the OPE is equivalent to the asymptotic small-mass
expansion of Feynman
integrals~\cite{GORISHNY1989633,Chetyrkin:1988zz,Chetyrkin:1988cu,Smirnov:1990rz,Smirnov}
and the knowledge of the perturbative condensates is necessary for
incorporating quartic quark-mass corrections. The first orders of
their diagramatic expansion read
\begin{equation}
  \label{eq:qq_exp}
    \langle \bar{q}q \rangle =\ \centeredgraphics[width=40pt]{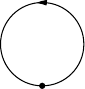}\ +\ \centeredgraphics[width=40pt]{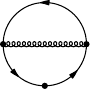}\ +\ \mathcal{O}(\alpha_s^2).
\end{equation}

One approach for bridging the gap between the low-energy and
high-energy regimes of QCD is renormalisation group optimised
perturbation theory
(RGOPT)~\cite{Kneur:2010ss,Kneur:2011vi,Kneur:2013coa}. This method
has been used to extrapolate from perturbative heavy-quark condensates
in the weakly-coupled regime to strongly coupled light-quark
condensates~\cite{Kneur:2015dda,Kneur:2020bph}. A crucial ingredient
in this analysis is the anomalous dimension of the condensate, which
has a well-known connection to the vacuum anomalous dimension
$\gamma_0$ \cite{Spiridonov:1988md}. Assuming $n_h$ quark flavours
with degenerate mass $m_q$ and $n_l$ massless quarks the relation
between the two reads
\begin{equation}
  \label{eq:vacuum_AD}
  \mu^2\frac{d}{d\mu^2} m_q \langle \bar{q} q \rangle = -4 m_q^4 \gamma_0.
\end{equation}
An RGOPT determination based on the four-loop heavy-quark
condensate~\cite{Maier:2019} and the five-loop vacuum anomalous
dimension~\cite{Baikov:2018nzi} found a notable numeric deviation from
lower-order results~\cite{Kneur:2015dda}, which could be rectified by
incorporating partial higher-order
contributions~\cite{Kneur:2020bph}. A completion of this analysis
requires the five-loop condensate and the six-loop anomalous
dimension.

In this note, we present the first five-loop QCD result including
non-vanishing quark masses.  We compute the perturbative heavy-quark
condensate at five loops and derive the vacuum anomalous dimension at
the same order. The computation of the anomalous dimension constitutes
a completely independent cross check of the results by Baikov
and Chetyrkin~\cite{Baikov:2018nzi}, which unlike other five-loop
anomalous dimensions in QCD
\cite{Baikov:2014qja,Baikov:2016tgj,Luthe:2016xec,Herzog:2017ohr,Baikov:2017ujl,Luthe:2017ttc,Luthe:2017ttg,Chetyrkin:2017bjc}
had not been confirmed by a second calculation so far.

\section{Calculational Strategy}
\label{sec:calc}

To compute the quark condensate at five loops, we first generate the
3451 contributing massive vacuum diagrams with the help of
\texttt{QGRAF}~\cite{Nogueira:1991ex}. In order to verify the gauge
independence of our results we insert the Feynman rules in a general
covariant gauge, expanding around Feynman gauge to linear order in the
gauge parameter.

\begin{figure}[htb]
  \centering
  \begin{tabular}{c@{\qquad}c@{\qquad}c@{\qquad}c}
    \includegraphics[width=0.1\textwidth]{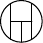}&
    \includegraphics[width=0.1\textwidth]{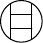}&
    \includegraphics[width=0.1\textwidth]{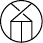}&
    \includegraphics[width=0.1\textwidth]{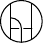}
  \end{tabular}
  \caption{The four distinct five-loop vacuum-type integral sectors with the maximum number of 12 propagators each. All our integrals can be mapped onto these or their sub-sectors.}
  \label{fig:tops}
\end{figure}

We simplify the resulting expressions using
\texttt{FORM}~\cite{Vermaseren:2000nd} code. In particular, we adopt
the \texttt{color}~\cite{vanRitbergen:1998pn} package to determine
colour factors. The resulting single-mass-scale scalar vacuum
integrals can be mapped onto 34 integral families, which correspond to
different mass colourings of the four five-loop sector templates
depicted in figure~\ref{fig:tops} as well as their subsectors that
form when contracting propagator lines to a point.

The scalar integrals can be reduced to a set of 142 master integrals
by exploiting integration-by-parts relations
\cite{Chetyrkin:1981qh}.\footnote{In~\cite{Maier:2024rng} the
  reduction lead to 156 different integrals. However, this number
  could be further reduced through a more comprehensive integral
  symmetrisation.} To this end, we employ
\texttt{crusher}~\cite{crusher}, a custom implementation of Laporta's
algorithm~\cite{Laporta:2001dd}, together with
\texttt{tinbox}~\cite{tinbox} for reduction over finite fields and
rational
reconstruction~\cite{Kauers:2008zz,Kant:2013vta,vonManteuffel:2014ixa,Peraro:2016wsq,Klappert:2019emp}. As
expected, the dependence on the gauge parameter cancels analytically
after the reduction to master integrals.

13 of the master integrals correspond to products of known integrals
with fewer loops. We compute the remaining master integrals via direct
numerical integration over one of the loop
momenta~\cite{Maier:2024rng}, validating the results against
low-precision expressions obtained with
FIESTA~\cite{Smirnov:2021rhf}. Details will be given
elsewhere. Finally, we obtain the quark condensate after
renormalisation in the \MS scheme.

To briefly illustrate the calculation, the initial five-loop expression has the form
\begin{equation}
  \label{eq:qq_initial}
  \langle \bar{q}q\rangle\Big\rvert_{\alpha_s^4} = \centeredgraphics{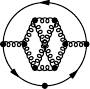} + (3450 \text{ further diagrams}).
\end{equation}
After inserting the Feynman rules and reducing the resulting scalar
integrals to basis integrals we obtain the form
\begin{equation}
  \label{eq:qq_red}
  \langle \bar{q}q\rangle\Big\rvert_{\alpha_s^4} = m_q^{3-10\epsilon}\,s(\epsilon)\,\centeredgraphics{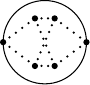} + (141 \text{ further basis integrals}),
\end{equation}
where $s(\epsilon)$ is a linear combination of colour factors with
Laurent series in the dimensional regulator $\epsilon = \frac{4-d}{2}$
as coefficients. $s(\epsilon)$ receives contributions from the diagram
explicitly shown in eq.~\eqref{eq:qq_initial} as well as from other
diagrams, for instance the ones where the gluons forming the inner
massless crossed box are replaced by light quark or ghost
propagators. With Euclidean scalar propagators $S(p) = \frac{1}{p^2 +
m^2}$, $m =0$ for the dotted and $m=1$ for the solid lines, and $[dl]
= e^{\epsilon \gamma_E}\pi^{-\frac{d}{2}} d^dl$ for the integration measure of each loop momentum
$l$, we obtain
\begin{equation*}
  \centeredgraphics{T5L10lF17} ={}
  \begin{aligned}[t]
    &-1.036927755143369926331365486457034168057080919501912811974192\,\epsilon^{-2}\\
    &-2.271517593918629612935996660135889091893839507169791336781934\,\epsilon^{-1}\\
    &-51.67721982612510097609793433215100054478372242(52) \\
    &-74.1136306549374748049570059276086520692992(10)\,\epsilon \\
    &-1534.58234483048602034962622627378789032210(13)\,\epsilon^2 \\
    &-997.92411520423778145046654481481444562108(29)\,\epsilon^3 \\
    &-37380.17801620860014113672827111290507436593(89)\,\epsilon^4 \\
    &+ \mathcal{O}(\epsilon^5),
  \end{aligned}
\end{equation*}
for the basis integral shown in eq.~\eqref{eq:qq_red}. The error in
the first two expansion orders is negligible. Note that for this
specific case, the coefficients of the $\epsilon$\/-expansion can be
inferred in analytic form in terms of (multiple) zeta values from the
known expansion of the massless 3-loop non-planar self-energy
insertion (see e.g.\ integral $M_{3,2}$ of~\cite{Lee:2011jt}). The
results agree, which we take as a strong check on our generic
integration method.

\section{Results}
\label{sec:res}

We present our results for the general case of a simple compact Lie
gauge group with a coupling constant $\alpha_s$. Fermions transform
according to the $N_F$-dimensional fundamental representation $(F)$ with the
quadratic Casimir operator $C_F$. The adjoint representation $(A)$ is
$N_A$-dimensional and we denote its quadratic Casimir operator by
$C_A$. We express higher-order group invariants in terms of
\begin{align}
  d^{44}_{R_1 R_2} ={}& \str(T_{R_1}^{a}T_{R_1}^{b}T_{R_1}^{c}T_{R_1}^{d}) \str(T_{R_2}^{a}T_{R_2}^{b}T_{R_2}^{c}T_{R_2}^{d}),
\end{align}
where $T_R^a$ denotes the $a$th generator in the representation $R \in \{F, A\}$,
and $\str$ a symmetrised trace~\cite{vanRitbergen:1998pn}. We adopt
the trace normalisation $\tr \left(T^a T^b\right) = T_F \delta^{a b}$.

\subsection{Quark Condensate and Vacuum Anomalous Dimension}
\label{sec:res_qq_gamma0}

With this notation, our result reads
\begin{align}
  \label{eq:qq}
  \langle \bar{q} q\rangle\Bigr|_{\mu = m_q} ={}& -\frac{N_F}{16 \pi^2} m_q^3 \sum_{i \geq 0} c_i \left(\frac{\alpha_s}{\pi}\right)^i,\displaybreak[0]\\
  c_0={}&4,\displaybreak[0]\\
  c_1={}&10\,C_F,\displaybreak[0]\\
  c_2={}&\bigg(-\frac{15}{2}-\frac{16}{3}\,\zeta(3)\bigg)\,C_F\,T_F\,n_l + \bigg(-\frac{125}{6}+\frac{56}{3}\,\zeta(3)\bigg)\,C_F\,T_F\,n_h\notag\\
&+\bigg(-\frac{139}{24}+\frac{8}{3}\,b_4+42\,\zeta(3)-\frac{22}{45}\,\pi^{4}\bigg)\,C_F^{2}\notag\\
&+\bigg(\frac{749}{24}-\frac{4}{3}\,b_4-\frac{49}{3}\,\zeta(3)+\frac{11}{45}\,\pi^{4}\bigg)\,C_F\,C_A,\displaybreak[0]\\
  c_3={}&\bigg(-\frac{163}{1458}+\frac{244}{27}\,\zeta(3)-\frac{7}{90}\,\pi^{4}\bigg)\,C_F\,T_F^{2}\,n_l^{2}\displaybreak[0]\notag\\
&+\bigg(\frac{10853}{729}-\frac{232}{27}\,\zeta(3)+\frac{1}{5}\,\pi^{4}\bigg)\,C_F\,T_F^{2}\,n_h\,n_l\displaybreak[0]\notag\\
&+\bigg(\frac{98977}{7290}+\frac{356}{135}\,\zeta(3)+\frac{1}{90}\,\pi^{4}\bigg)\,C_F\,T_F^{2}\,n_h^{2}\displaybreak[0]\notag\\
&+\bigg(-\frac{63883}{1728}-\frac{152}{27}\,b_4-\frac{32}{45}\,b_5-\frac{1151}{18}\,\zeta(3)+107\,\zeta(5)+\frac{5771}{3240}\,\pi^{4}\displaybreak[0]\notag\\
&-\frac{196}{135}\,\log(2)\,\pi^{4}\bigg)\,C_F^{2}\,T_F\,n_l+\bigg(-\frac{788273}{23328}+\frac{76}{27}\,b_4+\frac{16}{45}\,b_5-\frac{1073}{54}\,\zeta(3)\displaybreak[0]\notag\\
&-42\,\zeta(5)-\frac{751}{1296}\,\pi^{4}+\frac{98}{135}\,\log(2)\,\pi^{4}\bigg)\,C_F\,C_A\,T_F\,n_l+\bigg(-\frac{24793}{576}+\frac{152}{27}\,b_4\displaybreak[0]\notag\\
&+\frac{2269}{18}\,\zeta(3)-105\,\zeta(5)-\frac{4853}{3240}\,\pi^{4}\bigg)\,C_F^{2}\,T_F\,n_h+\bigg(-\frac{2824721}{23328}+\frac{500}{27}\,b_4\displaybreak[0]\notag\\
&+\frac{18619}{54}\,\zeta(3)-\frac{19}{2}\,\zeta(3)^{2}+79\,\zeta(5)-\frac{34189}{6480}\,\pi^{4}-\frac{1}{108}\,\pi^{6}\bigg)\,C_F\,C_A\,T_F\,n_h\displaybreak[0]\notag\\
&+\bigg(\frac{20789}{192}+44\,b_4+\frac{801}{2}\,\zeta(3)+2\,\zeta(3)^{2}+204\,\zeta(5)-\frac{981}{80}\,\pi^{4}+\frac{17}{756}\,\pi^{6}\bigg)C_F^{3}\displaybreak[0]\notag\\
&+\bigg(-\frac{123829}{1728}-\frac{200}{27}\,b_4+\frac{88}{45}\,b_5-\frac{4193}{72}\,\zeta(3)+\frac{249}{4}\,\zeta(3)^{2}-\frac{899}{2}\,\zeta(5)\displaybreak[0]\notag\\
&+\frac{8185}{2592}\,\pi^{4}-\frac{25}{1512}\,\pi^{6}+\frac{539}{135}\,\log(2)\,\pi^{4}\bigg)C_F^{2}\,C_A+\bigg(\frac{10491953}{93312} -\frac{197}{27}\,b_4\displaybreak[0]\notag\\
&-\frac{44}{45}\,b_5-\frac{2059}{216}\,\zeta(3)-\frac{137}{4}\,\zeta(3)^{2}+\frac{575}{4}\,\zeta(5)+\frac{18287}{12960}\,\pi^{4}-\frac{25}{3024}\,\pi^{6}\displaybreak[0]\notag\\
&-\frac{539}{270}\,\log(2)\,\pi^{4}\bigg)C_F\,C_A^{2},\displaybreak[0]\\
  c_4={}& - 11.3384529459116\,C_F\,T_F^3\,n_l^3
          - 14.603905334264\,C_F\,T_F^3\,n_l^2\,n_h\displaybreak[0]\notag\\
        & + 3.471093792654\,C_F\,T_F^3\,n_l\,n_h^2
          + 5.954796395928\,C_F\,T_F^3\,n_h^3\displaybreak[0]\notag\\
        & + 66.445670997293\,C_F^2\,T_F^2\,n_l^2
          + 141.501528239179\,C_F\,C_A\,T_F^2\,n_l^2\displaybreak[0]\notag\\
        & - 1.703206306289\,C_F^2\,T_F^2\,n_l\,n_h
          + 253.882002359631\,C_F\,C_A\,T_F^2\,n_l\,n_h\displaybreak[0]\notag\\
        & - 55.130662516989\,C_F^2\,T_F^2\,n_h^2
          + 88.024887370248\,C_F\,C_A\,T_F^2\,n_h^2\displaybreak[0]\notag\\
        & - 296.892299918978\,C_F^3\,T_F\,n_l
          - 261.509746457340\,C_F^2\,C_A\,T_F\,n_l\displaybreak[0]\notag\\
        & - 442.412410567868\,C_F\,C_A^2\,T_F\,n_l
          + 9.359717316091\,\frac{d^{44}_{FF}}{N_F}\,n_l\displaybreak[0]\notag\\
        & + 114.799301848235\,C_F^3\,T_F\,n_h
          - 364.277242381176\,C_F^2\,C_A\,T_F\,n_h\displaybreak[0]\notag\\
        & - 416.422797603641\,C_F\,C_A^2\,T_F\,n_h
          + 145.171194649991\,\frac{d^{44}_{FF}}{N_F}\,n_h\displaybreak[0]\notag\\
        & - 20.976740021973\,\frac{d^{44}_{FA}}{N_F}
          + 448.011364796477\,C_F^4\displaybreak[0]\notag\\
        & - 84.228640613689\,C_F^3\,C_A
          + 565.022325427743\,C_F^2\,C_A^2\displaybreak[0]\notag\\
        & + 301.083789093051\,C_F\,C_A^3
\end{align}
for the quark condensate and
\begin{align}
  \label{eq:gamma_0}
  \gamma_0 ={}& \frac{N_F}{16 \pi^2} \sum_{i\geq0} \gamma_{0i}\left(\frac{\alpha_s}{\pi}\right)^i,\displaybreak[0]\\
  \gamma_{00}={}&-1,\displaybreak[0]\\
  \gamma_{01}={}&-C_F,\displaybreak[0]\\
  \gamma_{02}={}&\frac{29}{8} C_F\,T_F\,n_h+\frac{5}{8} C_F\,T_F\,n_l+\bigg(\frac{131}{32}-3\,\zeta(3)\bigg)\,C_F^{2}+\bigg(-\frac{109}{32} +\frac{3}{2}\,\zeta(3)\bigg)\,C_F\,C_A,\displaybreak[0]\\
  \gamma_{03}={}&\bigg(\frac{341}{486}-\frac{2}{3}\,\zeta(3)\bigg)\,C_F\,T_F^{2}\,n_l^{2}+\bigg(\frac{233}{243}-\frac{4}{3}\,\zeta(3)\bigg)C_F\,T_F^{2}\,n_h\,n_l+\bigg(\frac{125}{486}-\frac{2}{3}\,\zeta(3)\bigg)C_F\,T_F^{2}\,n_h^{2}\notag\\
&+\bigg(\frac{281}{96}+\frac{1}{6}\,\zeta(3)-\frac{1}{36}\,\pi^{4}\bigg)\,C_F^{2}\,T_F\,n_l+\bigg(\frac{185}{96}+\frac{1}{6}\,\zeta(3)-\frac{1}{36}\,\pi^{4}\bigg)\,C_F^{2}\,T_F\,n_h\notag\\
&+\bigg(-\frac{661}{3888}+\frac{1}{4}\,\zeta(3)+\frac{1}{45}\,\pi^{4}\bigg)\,C_F\,C_A\,T_F\,n_l\notag\\
&+\bigg(\frac{39083}{3888}+\frac{23}{4}\,\zeta(3)-\frac{5}{2}\,\zeta(5)+\frac{1}{45}\,\pi^{4}\bigg)\,C_F\,C_A\,T_F\,n_h\notag\\
&+\bigg(-\frac{1471}{96}+\frac{3}{4}\,\zeta(3)+\frac{5}{2}\,\zeta(5)+\frac{1}{20}\,\pi^{4}\bigg)\,C_F^{3}\notag\\
&+\bigg(\frac{56}{3}-\frac{413}{24}\,\zeta(3)+\frac{45}{4}\,\zeta(5)+\frac{1}{180}\,\pi^{4}\bigg)\,C_F^{2}\,C_A\notag\\
&+\bigg(-\frac{121547}{15552}+\frac{235}{24}\,\zeta(3)-\frac{65}{8}\,\zeta(5)-\frac{11}{720}\,\pi^{4}\bigg)\,C_F\,C_A^{2},\displaybreak[0]\\
  \gamma_{04}={}& -0.000225895\,C_F\,T_F^3\,n_l^3
                  -0.111789\,C_F\,T_F^3\,n_h\,n_l^2
                  -0.2229\,C_F\,T_F^3\,n_h^2\,n_l\displaybreak[0]\notag\\
                  &-0.111337\,C_F\,T_F^3\,n_h^3
                  -0.624743\,C_F^2\,T_F^2\,n_l^2
                  -0.785432\,C_F\,C_A\,T_F^2\,n_l^2\displaybreak[0]\notag\\
                  &-0.122948\,C_F^2\,T_F^2\,n_h\,n_l
                  -18.6121\,C_F\,C_A\,T_F^2\,n_h\,n_l
                  +0.501795\,C_F^2\,T_F^2\,n_h^2\displaybreak[0]\notag\\
                  &-17.8267\,C_F\,C_A\,T_F^2\,n_h^2
                  +1.10308\,C_F^3\,T_F\,n_l
                  +7.50086\,C_F\,C_A^2\,T_F\,n_l\displaybreak[0]\notag\\
                  &-1.09626\,C_A\,C_F^2\,T_F\,n_l
                  -0.503324\frac{d^{44}_{FF}}{N_F}\,n_l
                  +6.90079\,C_F^3\,T_F\,n_h\displaybreak[0]\notag\\
                  &-5.07863\,C_F^2\,C_A\,T_F\,n_h
                  +50.5652\,C_F\,C_A^2\,T_F\,n_h
                  +14.1913\frac{d^{44}_{FF}}{N_F}\,n_h\displaybreak[0]\notag\\
                  &+15.0064\,C_F^4
                  -14.3487\,C_F\,C_A^3
                  +42.3735\,C_F^2\,C_A^2\displaybreak[0]\notag\\
                  &-50.6615\,C_F^3\,C_A
                  -36.2448\frac{d^{44}_{FA}}{N_F}
\end{align}
for the vacuum anomalous dimension. We have used the short-hand notation
\begin{align}
  b_4 ={}& 4!\Li_4\left(\tfrac{1}{2}\right) + \ln^2(2) \left( \ln^2(2) - \pi^2\right),\\
  b_5 ={}& 5!\Li_5\left(\tfrac{1}{2}\right) - \ln^3(2) \left( \ln^2(2) - \frac{5}{3}\pi^2\right).
\end{align}
$\zeta$ is the Riemann zeta function and $\Li$ the polylogarithm, viz.
\begin{align}
  \zeta(n) = \sum_{k=1}^\infty \frac{1}{k^n},\\
  \Li_n(z) = \sum_{k=1}^\infty \frac{z^k}{k^n}.
\end{align}
Since the analytic result for $\gamma_{04}$ is
known~\cite{Baikov:2018nzi}, we have refrained from listing all digits
obtained in our numerical calculation. Each numeric coefficient in our
full result for $\gamma_{04}$ agrees with the exact one to better than
$10^{-13}$, and we find exact agreement for the coefficients
$\gamma_{01}, \gamma_{02},
\gamma_{03}$.\footnote{In~\cite{Baikov:2018nzi}, a factor of
  $\frac{1}{16\pi^2}$ is missing in the expression for $\gamma_{00}$.}

\subsection{Specific Values and Series Convergence}
\label{sec:res_sun}

For a SU(N) gauge group, the group invariants are given by
\begin{align}
  &N_F = N, \ \qquad\qquad\qquad  C_F = \frac{N^2-1}{N}\*T_F, \ \qquad\qquad\qquad  C_A= 2NT_F, \\
  &d^{44}_{FF} = \frac{N^2-1}{6\*N^2}(N^4-6\*N^2+18)\*T_F^4, \qquad d^{44}_{FA} =\frac{N^2-1}{3}\*N\*(N^2+6)\*T_F^4.
\end{align}
 Specifically in QCD, we have $N_F = C_A = 3, C_F = \frac{4}{3},
d^{44}_{FF} = \frac{5}{12},d^{44}_{FA} = \frac{15}{2}$ with $T_F =
\frac{1}{2}$, and the analytic results read
\begin{align}
  c_1\Big\rvert_{\text{SU(3)}} ={}& \frac{40}{3},\displaybreak[0]\\
c_{2}\Big\rvert_{\text{SU(3)}} ={}&\frac{6185}{54}+\frac{28}{3}\,\zeta(3)-\frac{16}{27}\,b_{4}+\frac{44}{405}\,\pi^{4}-\left(5+\frac{32}{9}\,\zeta(3)\right)\,n_l+\left(-\frac{125}{9}+\frac{112}{9}\,\zeta(3)\right)\,n_h,\displaybreak[0]\\
c_{3}\Big\rvert_{\text{SU(3)}} ={}&\frac{9515801}{7776}+\frac{28315}{54}\,\zeta(3)-\frac{1844}{81}\,b_{4}+\frac{9151}{1944}\,\pi^{4}-\frac{176}{135}\,b_{5}-\frac{1699}{9}\,\zeta(5)-\frac{391}{2916}\,\pi^{6}\notag\\
&-\frac{1078}{405}\,\log(2)\,\pi^{4}-\frac{2005}{27}\,\zeta(3)^{2}+\left(-\frac{1171571}{11664}-\frac{7823}{81}\,\zeta(3)+\frac{152}{243}\,b_{4}\right.\notag\\
&\left.+\frac{12373}{29160}\,\pi^{4}+\frac{32}{405}\,b_{5}+\frac{100}{9}\,\zeta(5)+\frac{196}{1215}\,\log(2)\,\pi^{4}\right)\,n_l+\left(-\frac{3270995}{11664}\right.\notag\\
&\left.+\frac{64933}{81}\,\zeta(3)+\frac{10216}{243}\,b_{4}-\frac{69305}{5832}\,\pi^{4}+\frac{194}{3}\,\zeta(5)-\frac{1}{54}\,\pi^{6}-19\,\zeta(3)^{2}\right)\,n_h\notag\\
&+\left(-\frac{163}{4374}+\frac{244}{81}\,\zeta(3)-\frac{7}{270}\,\pi^{4}\right)\,n_l^{2}+\left(\frac{10853}{2187}-\frac{232}{81}\,\zeta(3)+\frac{1}{15}\,\pi^{4}\right)\,n_l\,n_h\notag\\
&+\left(\frac{98977}{21870}+\frac{356}{405}\,\zeta(3)+\frac{1}{270}\,\pi^{4}\right)\,n_h^{2},\displaybreak[0]\\
\gamma_{01}\Big\rvert_{\text{SU(3)}} ={}& -\frac{4}{3},\displaybreak[0]\\
\gamma_{02}\Big\rvert_{\text{SU(3)}}=&-\frac{457}{72}+\frac{2}{3}\,\zeta(3)+\frac{5}{12}\,n_l+\frac{29}{12}\,n_h,\displaybreak[0]\\
  \gamma_{03}\Big\rvert_{\text{SU(3)}}=&-\frac{39595}{1296}+\frac{55}{2}\,\zeta(3)-\frac{19}{540}\,\pi^{4}-\frac{1705}{54}\,\zeta(5)\notag\\
&+\left(\frac{4397}{1944}+\frac{35}{54}\,\zeta(3)+\frac{8}{405}\,\pi^{4}\right)\,n_l+\left(\frac{42413}{1944}+\frac{629}{54}\,\zeta(3)+\frac{8}{405}\,\pi^{4}-5\,\zeta(5)\right)\,n_h\notag\\
&+\left(\frac{341}{1458}-\frac{2}{9}\,\zeta(3)\right)\,n_l^{2}+\left(\frac{233}{729}-\frac{4}{9}\,\zeta(3)\right)\,n_l\,n_h+\left(\frac{125}{1458}-\frac{2}{9}\,\zeta(3)\right)\,n_h^{2},\displaybreak[0]\\
\gamma_{04}\Big\rvert_{\text{SU(3)}}=&-\frac{2113877}{9216}+\frac{661915}{1296}\,\zeta(3)-\frac{19081}{10368}\,\pi^{4}-\frac{27739}{27}\,\zeta(5)+\frac{113245}{326592}\,\pi^{6}+\frac{2677}{864}\,\zeta(3)^{2}\notag\\
&+\frac{80017}{288}\,\zeta(7)+\left(\frac{8080877}{186624}+\frac{21329}{2592}\,\zeta(3)+\frac{47303}{155520}\,\pi^{4}+\frac{491}{432}\,\zeta(5)-\frac{5155}{163296}\,\pi^{6}\right.\notag\\
&\left.-\frac{2165}{432}\,\zeta(3)^{2}-\frac{49}{16}\,\zeta(7)\right)\,n_l+\left(\frac{44749901}{186624}+\frac{461621}{2592}\,\zeta(3)-\frac{8003}{31104}\,\pi^{4}\right.\notag\\
&\left.-\frac{36583}{144}\,\zeta(5)+\frac{365}{40824}\,\pi^{6}+\frac{3269}{216}\,\zeta(3)^{2}+\frac{14875}{144}\,\zeta(7)\right)\,n_h+\left(-\frac{510101}{1119744}\right.\notag\\
  &\left.-\frac{4817}{1296}\,\zeta(3)+\frac{179}{25920}\,\pi^{4}+\frac{83}{27}\,\zeta(5)\right)\,n_l^{2}+\left(-\frac{5894549}{559872}-\frac{10001}{648}\,\zeta(3)+\frac{119}{3240}\,\pi^{4}\right.\notag\\
&   \left.+\frac{467}{54}\,\zeta(5)-\frac{5}{3024}\,\pi^{6}-\frac{3}{8}\,\zeta(3)^{2}\right)\,n_l\,n_h+\left(-\frac{11278997}{1119744}-\frac{15185}{1296}\,\zeta(3)\right.\notag\\
&   \left.+\frac{773}{25920}\,\pi^{4}+\frac{301}{54}\,\zeta(5)-\frac{5}{3024}\,\pi^{6}-\frac{3}{8}\,\zeta(3)^{2}\right)\,n_h^{2}+\left(\frac{373}{10368}+\frac{13}{648}\,\zeta(3)\right.\notag\\
&   \left.-\frac{1}{1620}\,\pi^{4}\right)\,n_l^{3}+\left(\frac{103}{1152}+\frac{13}{216}\,\zeta(3)-\frac{1}{540}\,\pi^{4}\right)\,n_l^{2}\,n_h+\left(\frac{245}{3456}+\frac{13}{216}\,\zeta(3)\right.\notag\\
&   \left.-\frac{1}{540}\,\pi^{4}\right)\,n_l\,n_h^{2}+\left(\frac{181}{10368}+\frac{13}{648}\,\zeta(3)-\frac{1}{1620}\,\pi^{4}\right)\,n_h^{3},
\end{align}
taking $\gamma_{04}$ from \cite{Baikov:2018nzi}. Numerically, we obtain for $n_h = 1$
\begin{align}
  \langle \bar{q} q\rangle\Bigr|_{\mu = m_q} ={}&
\begin{aligned}[t]
  -\frac{N_F}{16 \pi^2} m_q^3 \Biggl[&4 + 13.3333 \frac{\alpha_s}{\pi} + (132.723 - 9.27398\,n_l) \left(\frac{\alpha_s}{\pi}\right)^2\\
  &+(1320.23 - 134.592\,n_l + 1.05832\,n_l^2) \left(\frac{\alpha_s}{\pi}\right)^3\\
 &+ (17394.7 - 3448.7\,n_l + 168.599\,n_l^2 - 1.88974\,n_l^3) \left(\frac{\alpha_s}{\pi}\right)^4\Biggr],
\end{aligned}\\
\gamma_0 ={}&
\begin{aligned}[t]
  -\frac{N_F}{16 \pi^2} \Biggl[&1 + 1.33333 \frac{\alpha_s}{\pi} +  (3.12918 - 0.416667\,n_l) \left(\frac{\alpha_s}{\pi}\right)^2\\
  &+ (1.2853 - 4.75044\,n_l + 0.0332417\,n_l^2) \left(\frac{\alpha_s}{\pi}\right)^3\\
  &+ (-40.3549 - 24.6153\,n_l + 1.08173\,n_l^2 + 0.0000376492\,n_l^3) \left(\frac{\alpha_s}{\pi}\right)^4\Biggr].
\end{aligned}
\end{align}
The poor apparent perturbative convergence for the condensate is in
line with the expectation that individual factors in the Operator
Product Expansion in the \MS scheme suffer from large renormalon
contributions, which are expected to cancel in the complete
expansion. For a detailed discussion and proposed resolution
see~\cite{Beneke:2025hlg}.

\section{Conclusions}
\label{sec:conclusions}

We have presented the five-loop result for the heavy-quark
condensate. This is the first time that non-vanishing quark masses
have been incorporated in a QCD calculation at this order. We have
further derived the five-loop vacuum anomalous dimension,
independently confirming a result by Baikov and Chetyrkin
\cite{Baikov:2018nzi}.

Applications of our result include asymptotic small-mass expansions
and RGOPT determinations~\cite{Kneur:2015dda,Kneur:2020bph} of the
light-quark condensate. Furthermore, the present work is an important
milestone on the path towards future five-loop QCD calculations
involving massive vacuum diagrams, in particular decoupling of heavy
quarks, quark-mass determinations via sum rules, and Higgs boson
production and decay.

\section*{Acknowledgements}

The work of A.M.\ was supported in part by the Spanish Ministry of
Science and Innovation (PID2020-112965GB-I00,PID2023-146142NB-I00),
and by the Departament de Recerca i Universities from Generalitat de
Catalunya to the Grup de Recerca 00649 (Codi: 2021 SGR 00649).  This
project received funding from the European Union’s Horizon 2020
research and innovation programme under grant agreement No 824093.
Y.S.\ acknowledges support from ANID under FONDECYT project
No.\ 1231056 and Exploraci\'on Project No.\ 13250014.

\bibliographystyle{JHEP}
\bibliography{biblio}

\providecommand{\href}[2]{#2}\begingroup\raggedright\begin{thebibliography}{10}

\bibitem{Gell-Mann:1968hlm}
M.~Gell-Mann, R.~J. Oakes and B.~Renner, \emph{{Behavior of current divergences
  under SU(3) x SU(3)}},
  \href{http://dx.doi.org/10.1103/PhysRev.175.2195}{\emph{Phys. Rev.} {\bf 175}
  (1968) 2195--2199}.

\bibitem{Wilson:1969zs}
K.~G. Wilson, \emph{{Nonlagrangian models of current algebra}},
  \href{http://dx.doi.org/10.1103/PhysRev.179.1499}{\emph{Phys. Rev.} {\bf 179}
  (1969) 1499--1512}.

\bibitem{Politzer:1976tv}
H.~D. Politzer, \emph{{Effective Quark Masses in the Chiral Limit}},
  \href{http://dx.doi.org/10.1016/0550-3213(76)90405-3}{\emph{Nucl. Phys. B}
  {\bf 117} (1976) 397--406}.

\bibitem{GORISHNY1989633}
S.~Gorishny, \emph{Construction of operator expansions and effective theories
  in the ms scheme},
  \href{http://dx.doi.org/https://doi.org/10.1016/0550-3213(89)90622-6}{\emph{Nuclear
  Physics B} {\bf 319} (1989) 633--666}.

\bibitem{Chetyrkin:1988zz}
K.~G. Chetyrkin, \emph{{Operator Expansions in the Minimal Subtraction Scheme.
  1: The Gluing Method}},
  \href{http://dx.doi.org/10.1007/BF01017168}{\emph{Theor. Math. Phys.} {\bf
  75} (1988) 346--356}.

\bibitem{Chetyrkin:1988cu}
K.~G. Chetyrkin, \emph{{Operator Expansions in the Minimal Subtraction Scheme.
  2: Explicit Formulas for Coefficient Functions}},
  \href{http://dx.doi.org/10.1007/BF01028580}{\emph{Theor. Math. Phys.} {\bf
  76} (1988) 809--817}.

\bibitem{Smirnov:1990rz}
V.~A. Smirnov, \emph{{Asymptotic expansions in limits of large momenta and
  masses}}, \href{http://dx.doi.org/10.1007/BF02102092}{\emph{Commun. Math.
  Phys.} {\bf 134} (1990) 109--137}.

\bibitem{Smirnov}
V.~A. Smirnov, \emph{Applied Asymptotic Expansions in Momenta and Masses}.
\newblock Springer Berlin, Heidelberg, 2002,
  \href{http://dx.doi.org/10.1007/3-540-44574-9}{10.1007/3-540-44574-9}.

\bibitem{Kneur:2010ss}
J.~L. Kneur and A.~Neveu, \emph{{Renormalization Group Improved Optimized
  Perturbation Theory: Revisiting the Mass Gap of the O(2N) Gross-Neveu
  Model}}, \href{http://dx.doi.org/10.1103/PhysRevD.81.125012}{\emph{Phys. Rev.
  D} {\bf 81} (2010) 125012}, [\href{http://arxiv.org/abs/1004.4834}{{\tt
  1004.4834}}].

\bibitem{Kneur:2011vi}
J.~L. Kneur and A.~Neveu, \emph{{$Lambda^{\rm QCD}_{\rm MS}$ from
  Renormalization Group Optimized Perturbation}},
  \href{http://dx.doi.org/10.1103/PhysRevD.85.014005}{\emph{Phys. Rev. D} {\bf
  85} (2012) 014005}, [\href{http://arxiv.org/abs/1108.3501}{{\tt 1108.3501}}].

\bibitem{Kneur:2013coa}
J.-L. Kneur and A.~Neveu, \emph{{$\alpha_S$ from $F_\pi$ and Renormalization
  Group Optimized Perturbation Theory}},
  \href{http://dx.doi.org/10.1103/PhysRevD.88.074025}{\emph{Phys. Rev. D} {\bf
  88} (2013) 074025}, [\href{http://arxiv.org/abs/1305.6910}{{\tt 1305.6910}}].

\bibitem{Kneur:2015dda}
J.-L. Kneur and A.~Neveu, \emph{{Chiral condensate from renormalization group
  optimized perturbation}},
  \href{http://dx.doi.org/10.1103/PhysRevD.92.074027}{\emph{Phys. Rev. D} {\bf
  92} (2015) 074027}, [\href{http://arxiv.org/abs/1506.07506}{{\tt
  1506.07506}}].

\bibitem{Kneur:2020bph}
J.-L. Kneur and A.~Neveu, \emph{{Chiral condensate and spectral density at full
  five-loop and partial six-loop orders of renormalization group optimized
  perturbation theory}},
  \href{http://dx.doi.org/10.1103/PhysRevD.101.074009}{\emph{Phys. Rev. D} {\bf
  101} (2020) 074009}, [\href{http://arxiv.org/abs/2001.11670}{{\tt
  2001.11670}}].

\bibitem{Spiridonov:1988md}
V.~P. Spiridonov and K.~G. Chetyrkin, \emph{{Nonleading mass corrections and
  renormalization of the operators m psi-bar psi and g**2(mu nu)}}, {\emph{Sov.
  J. Nucl. Phys.} {\bf 47} (1988) 522--527}.

\bibitem{Maier:2019}
A.~Maier and P.~Marquard, ``The heavy-quark condensate at four loops.''
  unpublished.

\bibitem{Baikov:2018nzi}
P.~A. Baikov and K.~G. Chetyrkin, \emph{{QCD vacuum energy in 5 loops}},
  \href{http://dx.doi.org/10.22323/1.290.0025}{\emph{PoS} {\bf RADCOR2017}
  (2018) 025}.

\bibitem{Baikov:2014qja}
P.~A. Baikov, K.~G. Chetyrkin and J.~H. K\"uhn, \emph{{Quark Mass and Field
  Anomalous Dimensions to ${\cal O}(\alpha_s^5)$}},
  \href{http://dx.doi.org/10.1007/JHEP10(2014)076}{\emph{JHEP} {\bf 10} (2014)
  076}, [\href{http://arxiv.org/abs/1402.6611}{{\tt 1402.6611}}].

\bibitem{Baikov:2016tgj}
P.~A. Baikov, K.~G. Chetyrkin and J.~H. K\"uhn, \emph{{Five-Loop Running of the
  QCD coupling constant}},
  \href{http://dx.doi.org/10.1103/PhysRevLett.118.082002}{\emph{Phys. Rev.
  Lett.} {\bf 118} (2017) 082002}, [\href{http://arxiv.org/abs/1606.08659}{{\tt
  1606.08659}}].

\bibitem{Luthe:2016xec}
T.~Luthe, A.~Maier, P.~Marquard and Y.~Schr\"oder, \emph{{Five-loop quark mass
  and field anomalous dimensions for a general gauge group}},
  \href{http://dx.doi.org/10.1007/JHEP01(2017)081}{\emph{JHEP} {\bf 01} (2017)
  081}, [\href{http://arxiv.org/abs/1612.05512}{{\tt 1612.05512}}].

\bibitem{Herzog:2017ohr}
F.~Herzog, B.~Ruijl, T.~Ueda, J.~A.~M. Vermaseren and A.~Vogt, \emph{{The
  five-loop beta function of Yang-Mills theory with fermions}},
  \href{http://dx.doi.org/10.1007/JHEP02(2017)090}{\emph{JHEP} {\bf 02} (2017)
  090}, [\href{http://arxiv.org/abs/1701.01404}{{\tt 1701.01404}}].

\bibitem{Baikov:2017ujl}
P.~A. Baikov, K.~G. Chetyrkin and J.~H. K\"uhn, \emph{{Five-loop fermion
  anomalous dimension for a general gauge group from four-loop massless
  propagators}}, \href{http://dx.doi.org/10.1007/JHEP04(2017)119}{\emph{JHEP}
  {\bf 04} (2017) 119}, [\href{http://arxiv.org/abs/1702.01458}{{\tt
  1702.01458}}].

\bibitem{Luthe:2017ttc}
T.~Luthe, A.~Maier, P.~Marquard and Y.~Schröder, \emph{{Complete
  renormalization of QCD at five loops}},
  \href{http://dx.doi.org/10.1007/JHEP03(2017)020}{\emph{JHEP} {\bf 03} (2017)
  020}, [\href{http://arxiv.org/abs/1701.07068}{{\tt 1701.07068}}].

\bibitem{Luthe:2017ttg}
T.~Luthe, A.~Maier, P.~Marquard and Y.~Schröder, \emph{{The five-loop Beta
  function for a general gauge group and anomalous dimensions beyond Feynman
  gauge}}, \href{http://dx.doi.org/10.1007/JHEP10(2017)166}{\emph{JHEP} {\bf
  10} (2017) 166}, [\href{http://arxiv.org/abs/1709.07718}{{\tt 1709.07718}}].

\bibitem{Chetyrkin:2017bjc}
K.~G. Chetyrkin, G.~Falcioni, F.~Herzog and J.~A.~M. Vermaseren,
  \emph{{Five-loop renormalisation of QCD in covariant gauges}},
  \href{http://dx.doi.org/10.1007/JHEP10(2017)179}{\emph{JHEP} {\bf 10} (2017)
  179}, [\href{http://arxiv.org/abs/1709.08541}{{\tt 1709.08541}}]. [Addendum:
  JHEP 12, 006 (2017)].

\bibitem{Nogueira:1991ex}
P.~Nogueira, \emph{{Automatic Feynman graph generation}},
  \href{http://dx.doi.org/10.1006/jcph.1993.1074}{\emph{J. Comput. Phys.} {\bf
  105} (1993) 279--289}.

\bibitem{Vermaseren:2000nd}
J.~A.~M. Vermaseren, \emph{{New features of FORM}},
  \href{http://arxiv.org/abs/math-ph/0010025}{{\tt math-ph/0010025}}.

\bibitem{vanRitbergen:1998pn}
T.~van Ritbergen, A.~N. Schellekens and J.~A.~M. Vermaseren, \emph{{Group
  theory factors for Feynman diagrams}},
  \href{http://dx.doi.org/10.1142/S0217751X99000038}{\emph{Int. J. Mod. Phys.
  A} {\bf 14} (1999) 41--96}, [\href{http://arxiv.org/abs/hep-ph/9802376}{{\tt
  hep-ph/9802376}}].

\bibitem{Chetyrkin:1981qh}
K.~G. Chetyrkin and F.~V. Tkachov, \emph{{Integration by Parts: The Algorithm
  to Calculate beta Functions in 4 Loops}},
  \href{http://dx.doi.org/10.1016/0550-3213(81)90199-1}{\emph{Nucl. Phys.} {\bf
  B192} (1981) 159--204}.

\bibitem{Maier:2024rng}
A.~Maier, P.~Marquard and Y.~Schr\"oder, \emph{{Towards QCD at five loops}},
  \href{http://dx.doi.org/10.22323/1.467.0084}{\emph{PoS} {\bf LL2024} (2024)
  084}, [\href{http://arxiv.org/abs/2407.16385}{{\tt 2407.16385}}].

\bibitem{crusher}
P.~Marquard and D.~Seidel, ``The {IBP} package \texttt{Crusher}.'' unpublished.

\bibitem{Laporta:2001dd}
S.~Laporta, \emph{{High precision calculation of multiloop Feynman integrals by
  difference equations}},
  \href{http://dx.doi.org/10.1016/S0217-751X(00)00215-7,
  10.1142/S0217751X00002157}{\emph{Int. J. Mod. Phys.} {\bf A15} (2000)
  5087--5159}, [\href{http://arxiv.org/abs/hep-ph/0102033}{{\tt
  hep-ph/0102033}}].

\bibitem{tinbox}
A.~Maier and P.~Marquard, ``\texttt{tinbox}, a finite field solver.''
  unpublished.

\bibitem{Kauers:2008zz}
M.~Kauers, \emph{{Fast solvers for dense linear systems}},
  \href{http://dx.doi.org/10.1016/j.nuclphysbps.2008.09.111}{\emph{Nucl. Phys.
  B Proc. Suppl.} {\bf 183} (2008) 245--250}.

\bibitem{Kant:2013vta}
P.~Kant, \emph{{Finding Linear Dependencies in Integration-By-Parts Equations:
  A Monte Carlo Approach}},
  \href{http://dx.doi.org/10.1016/j.cpc.2014.01.017}{\emph{Comput. Phys.
  Commun.} {\bf 185} (2014) 1473--1476},
  [\href{http://arxiv.org/abs/1309.7287}{{\tt 1309.7287}}].

\bibitem{vonManteuffel:2014ixa}
A.~von Manteuffel and R.~M. Schabinger, \emph{{A novel approach to integration
  by parts reduction}},
  \href{http://dx.doi.org/10.1016/j.physletb.2015.03.029}{\emph{Phys. Lett. B}
  {\bf 744} (2015) 101--104}, [\href{http://arxiv.org/abs/1406.4513}{{\tt
  1406.4513}}].

\bibitem{Peraro:2016wsq}
T.~Peraro, \emph{{Scattering amplitudes over finite fields and multivariate
  functional reconstruction}},
  \href{http://dx.doi.org/10.1007/JHEP12(2016)030}{\emph{JHEP} {\bf 12} (2016)
  030}, [\href{http://arxiv.org/abs/1608.01902}{{\tt 1608.01902}}].

\bibitem{Klappert:2019emp}
J.~Klappert and F.~Lange, \emph{{Reconstructing rational functions with
  FireFly}}, \href{http://dx.doi.org/10.1016/j.cpc.2019.106951}{\emph{Comput.
  Phys. Commun.} {\bf 247} (2020) 106951},
  [\href{http://arxiv.org/abs/1904.00009}{{\tt 1904.00009}}].

\bibitem{Smirnov:2021rhf}
A.~V. Smirnov, N.~D. Shapurov and L.~I. Vysotsky, \emph{{FIESTA5: Numerical
  high-performance Feynman integral evaluation}},
  \href{http://dx.doi.org/10.1016/j.cpc.2022.108386}{\emph{Comput. Phys.
  Commun.} {\bf 277} (2022) 108386},
  [\href{http://arxiv.org/abs/2110.11660}{{\tt 2110.11660}}].

\bibitem{Lee:2011jt}
R.~N. Lee, A.~V. Smirnov and V.~A. Smirnov, \emph{{Master Integrals for
  Four-Loop Massless Propagators up to Transcendentality Weight Twelve}},
  \href{http://dx.doi.org/10.1016/j.nuclphysb.2011.11.005}{\emph{Nucl. Phys. B}
  {\bf 856} (2012) 95--110}, [\href{http://arxiv.org/abs/1108.0732}{{\tt
  1108.0732}}].

\bibitem{Beneke:2025hlg}
M.~Beneke and H.~Takaura, \emph{{Gradient-flowed operator product expansion
  without IR renormalons}},
  \href{http://dx.doi.org/10.1007/JHEP03(2026)033}{\emph{JHEP} {\bf 03} (2026)
  033}, [\href{http://arxiv.org/abs/2510.12193}{{\tt 2510.12193}}].

\end{thebibliography}\endgroup

\end{document}